\documentclass[conference]{IEEEtran}

\usepackage{latexsym}
\usepackage{amsmath}
\usepackage{amsfonts}
\usepackage{tikz}
\usepackage{subfigure}

\usepackage{url}

\newcommand{\remove}[1]{}

\newtheorem{definition}{Definition}[section]

\newtheorem{example}{Example} 

\newcommand{\comment}[1]{}                           						
\newcommand{\ignore}[1]{\comment{#1}}                           				

\begin{document}
\title{Secure and Policy-Private Resource Sharing in an Online Social Network}

%
%
\author{\IEEEauthorblockN{Stefano Braghin}
\IEEEauthorblockA{DICOM\\
University of Insubria
}
\and
\IEEEauthorblockN{Vincenzo Iovino}
\IEEEauthorblockA{DIA\\
University of Salerno
}
\and
\IEEEauthorblockN{Giuseppe Persiano}
\IEEEauthorblockA{DIA\\
University of Salerno
}
\and
\IEEEauthorblockN{Alberto Trombetta}
\IEEEauthorblockA{DICOM\\
University of Insubria
}
}

\maketitle

\begin{abstract}
Providing functionalities that allow online social network users to manage in a secure and private way the publication of their information and/or resources is a relevant and far from trivial topic that has been under scrutiny from various research communities. In this work, we provide a framework that allows users to define highly expressive access policies to their resources in a way that the enforcement does not require the intervention of a (trusted or not) third party. 
This is made possible by the deployment of a newly defined cryptographic primitives
that provides - among other things - efficient access revocation and access policy privacy.
Finally, we provide an implementation of our framework as a Facebook application, proving the feasibility of our approach.
\end{abstract}

\newcommand{\primName}{Distance-Based Revokable Attribute Encryption}
\newcommand{\primAcr}{{\sf DBRA}}
\newcommand{\kg}{\primAcr{\sf .KG}}
\newcommand{\HKg}{{\sf HVE.KG}}
\newcommand{\IKg}{{\sf HIBE.KG}}
\newcommand{\Kg}{{\sf KG}}
\newcommand{\enc}{\primAcr{\sf .Enc}}
\newcommand{\HEnc}{{\sf HVE.Enc}}
\newcommand{\IEnc}{{\sf HIBE.Enc}}
\newcommand{\HDec}{{\sf HVE.Dec}}
\newcommand{\IDec}{{\sf HIBE.Dec}}
\newcommand{\HDerive}{{\sf HVE.Derive}}
\newcommand{\IDerive}{{\sf HIBE.Derive}}
\newcommand{\IDelegate}{\sf HIBE.Delegate}
\newcommand{\IRevoke}{\sf HIBE.Revoke}
\newcommand{\Enc}{{\sf Enc}}
\newcommand{\dec}{\primAcr{\sf .Dec}}
\newcommand{\Dec}{{\sf Dec}}
\newcommand{\derive}{\primAcr{\sf .Derive}}
\newcommand{\Derive}{{\sf Derive}}
\newcommand{\delegate}{\primAcr{\sf .Delegate}}
\newcommand{\Delegate}{{\sf Delegate}}
\newcommand{\revoke}{\primAcr{\sf .Revoke}}
\newcommand{\Revoke}{\sf Revoke}
\newcommand{\match}{{\sf Match}}
\newcommand{\prefix}{{\sf prefix}}
\newcommand{\Match}{{\sf Match}}
\newcommand{\HMatch}{{\sf HVE.Match}}
\newcommand{\pk}{{\tt pk}}
\newcommand{\hpk}{{\tt HVE.pk}}
\newcommand{\ipk}{{\tt HIBE.pk}}
\newcommand{\sk}{{\tt sk}}
\newcommand{\msk}{{\tt msk}}
\newcommand{\hmsk}{{\tt HVE.msk}}
\newcommand{\imsk}{{\tt HIBE.msk}}
\newcommand{\ct}{{\tt ct}}
\newcommand{\key}{{\tt key}}

\newcommand{\ske}{{\sf SKE}}
\newcommand{\skg}{{\ske}{\sf .KG}}
\newcommand{\senc}{{\ske}{\sf .Enc}}
\newcommand{\e}{{\ensuremath e}}

\newcommand{\x}{{\vec x}}
\newcommand{\y}{{\vec y}}

\newcommand{\pol}{{\tt Pol}}
\newcommand{\sdec}{{\ske}{\sf .Dec}}
\newcommand{\z}{{\vec z}}

\section{Introduction}
\label{sec:intro}
Online social networks (OSNs from now on) are nowadays used by hundreds of millions of users as the primary way of interacting and sharing of information and digital resources, as examples
such as Facebook, Flickr, LinkedIn and many others OSNs testify.
As the widespread adoption of OSNs increases and diversifies into various contexts (such as corporate OSNs),
the need for mechanisms protecting and regulating the publishing of users' sensitive resources is becoming paramount. 
The totality of OSNs allows users to specify -- usually in a rather limited way -- policies protecting their privacy and the confidentiality of their
sensitive data/resources. In this way, access to such data is restricted to the users that satisfy such policies.
Usually, the enforcement of users' specified policies is delegated to the central authority that manages the entire OSN,
and this state of affairs assumes -- of course -- a large degree of trust in such central authority.
This may be not acceptable when the shared resources are highly sensitive 
(as, for example, in the case of healthcare-related data). 
A (quite radical) solution
would delegate the entire management (storage and access control) 
of resources to their legitimate owner. This is clearly not feasible in a real world setting.
\ignore{
Rather, it would be desirable that the owner is the sole responsible for the enforcement of access policies to her/his resources and it does so 
by appropriately encrypting the resources and by releasing 
restricted decryption keys to potential users according to their credentials.
Having done that, the encrypted resource can be stored on a 
possibly untrusted server.
}

In this work, we propose a framework for publishing resources 
and for establishing social links
over an OSN 
that gives the owner of the resource a fine-grained control on who can access the resources without
having to trust the manager of the OSN; in our framework, 
the task of the OSN manager is reduced to that of 
providing reliable storage of the resources.
Specifically, in our framework user can establish a social relationship with other users and 
for each such relationship can specify the type of resources that can be accessed. In addition,
for each published resource the owner can specify an access policy specifying the type
of relationship needed for a user to access it.
Enforcement of access policies is guaranteed by means of novel cryptographic 
primitive, {\em \primName} (\primAcr\ in short, see Section~\ref{sec:cryp}), 
for which we give an efficient implementation based
on the Hidden Vector Encryption of ~\cite{DBLP:conf/pairing/IovinoP08} and
the Hierarchical Identity-Based Encryption of ~\cite{DBLP:conf/eurocrypt/BonehBG05}.
Roughly speaking, in the implementation of the framework based on 
\primAcr, the establishment of a link involves the transfer of a set of restricted decryption keys
that depend on the type of resources that can be accessed. On the other hand, publishing a resource
$res$ involves publishing an appropriately encrypted version of $res$. 
We stress that restricted decryption keys are transferred only when a new link is established
(or removed).


\paragraph{Access policies and key propagation}
In our framework, a user can express access policies that are closely tied 
to the OSN graph that is modeled
as a directed graph in which nodes represent the users of the OSN and 
edges represent relationships among users. 
For example, an access control policy may state that a user directly connected to the resource owner
is able to access the resource $res$ (given that she/he possibly satisfies other additional conditions).
In this case, our framework guarantees that $res$ is encrypted in such a way 
that keys held by all directly connected users are sufficient for decryption.
Notice that this holds even for users that 
have \emph{not} made any explicit request for accessing $res$ and for keys that were transferred
even before $res$ was published. 
Our framework also allows to express policies in terms of the distance in the OSN. 
For example, an access policy might specify a maximum
distance at which a resource can be accessed.
For such policies our framework provides a way for users to propagate decryption keys
to their neighbors until the limit distance from the owner, as specified by the access policy, 
has been reached.
\paragraph{Key management}
In our framework a user needs to generate 
only one key (the master secret key \msk) during enrollment. The master secret key is then used
to derive the appropriate keys that are transferred whenever a new link is created. 
Publishing a resource does not involve generating new keys and thus the number of keys that a user
need to manage depends only on the number of incoming links in the OSN and not on the number 
of resources he or his neighbors have published. Similarly, the size of the master secret key of a 
user does not depend on the number of resources published or on the number of outgoing links.
New  resources and links can be added without affecting already published resources and already 
established links.

\paragraph{Non-Interactive access to resources.}
We also stress that access to a resource by a user that has the appropriate keys 
can be achieved without interacting with the resource owner. 
Rather the users downloads the encrypted version of the resource from the OSN and uses keys obtained
during the establishment of the link with the owner of the resource to decrypt it.

\paragraph{Privacy of access policies}
In several settings, the access policy associated with a resource is itself sensitive information.
In our framework the access policy for a resource is determined by the owner upon publishing 
the resource and affects the way the resource is encrypted. \primAcr\ not only guarantees that 
the resource cannot be accessed without an appropriate key but also that the associated policy 
is kept confidential. All that a user can do is to try to decrypt the resource using the keys that 
he has received during the establishment of the social link with the owner of the resources.
If decryption fails then the user does not have any information on why the decryption has failed
(e.g., which credential was missing).

\paragraph{Fast revocation}
An OSN is for its own nature extremely dynamic and thus it is expected that relationship 
links might be removed. One way to deal with such a case would be to have the user re-enroll,
re-encrypt all resources and generate new keys for all of his neighbors. 
We stress that in case a link is revoked it is necessary to modify each encrypted resource,
for otherwise nothing would prevent the removed user to use the old keys to access the 
unmodified resource.  Our revocation procedure does so without having to completely recompute
the encryption. In turn, if encryptions of resources are modified then old keys do not
work anymore and need to be updated. Again, our revocation procedure does so
without having to generate keys from scratch.
Obviously, revocation is not retroactive and thus a user retains all resources that he 
has decrypted before the link was revoked.

\paragraph{Flexible OSN}
In the case a user missing a relationship link with the resource 
owner wishes to access the corresponding resource, it may ask for the creation of a link. After the link
has been created, the decryption key is passed to it, provided -- of course -- that it satisfies the 
other conditions expressed in the access policy.
Since the OSN may be very dynamic, with creation and deletion of friendship links, our access control 
mechanisms offer the possibility to update an access policy, by allowing a resource owner to 
revoke the access to a set of previously entitled users.

We do not assume that all links and all resources published by a user
belong to the same social (sub-)network but rather we allow each user 
to specify for each resource and for each
relationship a set of attributes 
(that identifies the social sub-network to which they belong).
Specifically, resources are labeled with pairs $(\x,d)$ 
where $\x$ is a vector of some pre-specified length $\ell$ encoding (conditions over) resources' attributes
and $d$ is an integer encoding distance.
Links instead are labeled with pairs $(\y,d')$ where $\y$ is also a vector of length $\ell$ 
and $d'$ is an integer.
If user Alice publishes a resource specifying $(\x,d)$, then 
user Bob sharing a link of type $(\y,d')$ with Alice can 
access the resource if and only if $\match((\x,d),(\y,d'))={\sf true}$.
The $\match$ predicate is true iff and only if 
vectors $\x$ and $\y$ agree in all positions in which $\y$ is not $\star$
and $d'\leq d$.


We now introduce our motivating scenario, describing how to use our ABE scheme to codify social relationships.  

\begin{example}
\label{ex:Alice}
Alice uses an OSN to keep in touch with people she knows and she decides to classify his published resources according to the following $2$ categories:
\emph{Friends} and \emph{Coworkers}.
Alice has friends from high school, college and the music club or neighbors;
similarly co-workers can be partitioned according to the $3$ 
different projects Alice is working on. 
Thus a resource is labeled  with a vector of length $2$ over 
$\{0,1,2,3,4\}$. Notice that since Alice is only involved in $3$ projects,
no resource will ever be labeled with a vector having $4$ as second component.
In addition for each resource Alice specifies a maximum distance at which
the resource can be accessed.
Obviously any such classification is likely to be partial and to change over
time; as we shall see, our system is flexible enough to allow 
Alice to dynamically and efficiently 
change the classification of her links and resources.

\end{example}


%

\section{Our framework}
\label{sec:framework}

We assume that each user $u$ owns a set $R_{u}=\{r_{1},r_{2},\ldots,r_{n}\}$ of resources. Each resource is denoted by a name $res$ and by a tuple of pairs
(attribute, value) $((att_{1}, val_{1}), (att_{2}, val_{2})\ldots,(att_{m},val_{m}))$,
where $att_{i}$, $1\leq i \leq m$ are attribute names. Attributes describe relevant features (in terms of access control purposes) of the
resource and take corresponding values $val_{1}, val_{2},\ldots,val_{m}$. We represent a resource with the expression $res(a_{1}, a_{2},\ldots,a_{m})$.
Further, we assume that each user is in charge of her/his own set of resources.
Attributes not only represent features of of resources to be accessed, but may also store information about their owners, for example who they are or where they work. 
We refer to such sets of attributes as credentials.
Given a resource (or -- more generally -- a set of resources), a corresponding access policy expresses under what conditions
a requester may gain access to the stated resource. More specifically, an access policy for a resource $res$ specifies the predicates that attributes
and corresponding values stored in resources/credentials owned by a user $u$ have to satisfy in order for $u$ to access resource $res$. 

\begin{example}
Now suppose Alice wants to publish an announcement regarding a concert to
all of her friends from the music club; then she can specify attribute 
vector $\langle3,0\rangle$. Here 
the first attribute (that is, $3$) specifies the category
of friends to which it is addressed ($3$ stands for friends from
the music club) and 
the second attribute specifies the category of co-workers 
(in this case Alice does not want co-workers to receive the announcement about the concert from her and thus specifies $0$). 
In addition Alice can specify a maximum distance $d$ she wants are announcement to be propagated. For example,
$(\langle3,0\rangle,2)$ makes the concert announcement accessible
to Alice's friends from the music club and to their friends.

\end{example}


\subsection{The policy language}	
\label{sec:lang}
We express access control policies by means of \emph{access rules} specified by resource owners. Access rules express under what conditions -- which we divide into
conditions about distance and  about credential attributes -- are to be satisfied a requesting user gain access to a stated resource.
Such access rules and policies are specified by the resource's owner.

We introduce the (fairly standard) language we will use to express access policies over resources in the OSN, starting with the definition of distance condition and
attribute-based condition.
Then, we introduce the precise definition of access rule and access policy.

An \emph{attribute condition} is written as $u.res.att pred val$, where $att$ is the name of an attribute belonging to resource $res$ (we omit the user and/or resource
name when there is no ambiguity in referring to the right attribute), $pred$ is a predicate belonging to $\{=, \leq, <, >, \geq,\not =\}$ and $val$ is valid value for the attribute
$att$.
a \emph{distance condition} is written as $dist(u,d)$ and it is verified in the case that there is a path between user $u$ (usually requesting access to resource $res$)
 and the user stating such condition having
length less or equal than $d$.
Given (attribute or distance) conditions $cond_{1},cond_{2},\ldots,cond_{k}$ and a resource $res$, a \emph{access rule} $ar^{res}$ for the resource $res$ is
written as  
$res\leftarrow \langle cond_{1},cond_{2},\ldots,cond_{k}\rangle$. The meaning of such expression is that, in order to access resource $res$,
the requester has to possess resources satisfying all the conditions
$cond_{1}, cond_{2},\ldots,cond_{u}$.
When the resource $res$ is clear from the context, we will omit it, writing the conditions' list only: $ar^{res}=cond_{1},cond_{2},\ldots,cond_{u}$. 
An \emph{access policy} $ap^{res}$ for the resource $res$ is defined as the set
$\{\langle ar^{res}_{1}\rangle ,\langle ar^{res}_{2}\rangle,\ldots,\langle ar^{res}_{v}\rangle\}$, where $ar^{res}_{i}$, $1\leq i\leq v$ are access rules. A resource $res$, protected by a corresponding
access policy $ap_{res}$, can be accessed by requester $u$ in the case that $u$ satisfies at least one of the access rules in $ap^{res}$.    
Such access control language allows for quite expressive access policies defined over resources owned by users of the OSN.

\begin{example}
Using the above presented policy language, the access policy for Alice's announcement $ann$ and stated 
by Alice herself can be written as 

\begin{equation*}
ar^{ann}_{1} = FriendType=``music club'', dist(u,2)
\end{equation*}

In addition, Alice wants to disclose such announcement to her college friends as well, and this can be expressed as

\begin{equation*}
ar^{ann}_{2} = FriendType=``college'', dist(u,1)
\end{equation*}

Overall, the access policy is thus expressed as

\begin{eqnarray*}
ap^{ann} &=&\{\langle FriendType=``music club'', dist(u,2)\rangle\\
&& \langle FriendType=``college'', dist(u,1)\rangle\}
\end{eqnarray*}
%
%
Note that the condition $dist(u,2)$ means that users accessing have to be at most at distance $1$ from Alice in the OSN graph, that is they have to
be friends with Alice, or friends with Alice's friends.
\end{example}


\subsection{Resource publication and access}
\label{sec:pubacc}
Once a user $u$ has defined the appropriate access control policy $pol^{res}$ for resource $res$, he computes the encryption $\hat{r}$ of the resource $res$ using the
\primAcr\ scheme, described in Section \ref{sec:cryp}, and
(apart the resource $res$, of course) the policy
$pol^{res}$. Then, the user $u$ publishes $\hat{r}$ on the (possibly untrusted) OSN repository. 

After having published the encrypted version of the resource $res$, the user $u$ propagates -- according to the access policy $pol^{res}$ -- to his friends 
$u_{1},u_{2},\ldots,u_{l}$ that satisfy the access control policy $pol^{res}$ (and only to them) the corresponding decryption keys
$k_{u_{1}},k_{u_{2}},\ldots,k_{u_{l}}$.

Once published on the OSN repository, the metadata related to the encrypted resource $\hat{res}$ (including, for example, the owner's identifier) can be searched
by all the OSN's users. In this way, once a OSN user $u'$ -- without any connection with user $u$ -- is willing to access resource $res$, $u'$ sends a
link creation request to $u$. Such request contains the attributes of $u'$ as well. Upon checking whether the attributes of $u'$ satisfy the access policy
$pol^{res}$, $u$ decides upon the creation of a friendship link with $u'$. In the positive case, $u$ computes a decryption key $k_{u'}$ and sends it to $u'$.
Of course, a user has as many decryption keys as sensitive resource it is entitled to access. We call such set of decryption keys the \emph{key ring} of user
$u$.

\begin{example}
Alice can specify attributes also for relationships by giving a pair
$(\y,d)$ where $\y$ is a vector of length $2$ over 
$\tilde\Sigma=\{0,1,2,3,4,\star\}$.
For example,
Alice's link to Bob, a senior colleague of Alice's working on $Project1$,
could be be labeled with the pair $(\langle 0,1\rangle,1)$.  
The distance in this case can be used to assign weights to links.
For example, the link to Carol, a junior colleague of Alice's working
on Project1, could be labeled with the pair $(\langle 0,1\rangle,2)$.  
In this way a document addressed to senior personnel of $Project1$ 
could be labeled with $(\langle 0,1\rangle,1)$.
In classifying links, a user is also allowed to use wild cards. 
For example, Alice's supervisor David will be linked to Alice with a link
labeled $(\langle 0,\star\rangle,1)$ which gives David access
to documents from all projects Alice is working on.

As we have seen, 
resources are tagged with an attribute vector and a distance 
encoding the set of users and the distance in the social 
up to which owner allows the resource to be accessed. 
Going back to our previous example of Alice publishing a 
concert announcement with attribute 
$\langle3,0\rangle,2$, we make two remarks.
First of all, the distance restriction does not mean that the 
announcement cannot be made available at distance greater than $2$ 
from Alice. 
Indeed, a friend of Alice's can read the announcement and 
re-post it herself.
The distance restriction only guarantees that if the announcement is 
read at distance greater than $2$ from Alice, then it cannot be linked 
to Alice.
As a second remark, 
in our OSN, the attributes to links and to resources are not to be taken
as recommendations. For example, by tagging the 
concert announcement with  
attribute $\langle 3,0\rangle$,
Alice does not mean that co-workers should not 
look at the announcement and that she relies on the the co-workers' 
honesty in not looking at the announcement 
(experience shows that such a labeling  would be too strong of a temptation
to resist for most people). 
Rather the resource will not be made available to co-workers and 
this will not be enforced by the central manager of the OSN but rather 
we will provide Alice with a publishing procedure that will 
cryptographically
enforce Alice's decision of the set of users with which we wants to share 
the announcement.

\end{example}

\section{Cryptographic notions}
\label{sec:cryp} 
\label{primacr}
In this section we present a new cryptographic primitive that 
we call \primName\ (\primAcr, in short)
that will constitute the main technical tool we use to enforce
privacy policy in our OSN.
In Section \ref{sec:primitives} we show how to construct \primAcr\ and in section \ref{sec:enfo}
we show how \primAcr\ is employed in our construction.

A \primAcr\ scheme consists of six algorithms
$(\kg,\enc,\dec,\derive,$ $\delegate,\revoke)$.
The key generation algorithm $\kg$ returns an {\em encryption key} $\pk$ 
and a {\em master secret key} $\msk$.  
The encryption algorithm $\enc$ takes as input 
the encryption key $\pk$, a plaintext $M$ and 
{\em ciphertext type} in the form of a pair
$(\x,d)$ consisting of an {\em attribute vector} $\x$ of length $\ell$ 
over a fixed alphabet $\Sigma$ and a integer {\em distance} $d$.
The master secret key $\msk$ can be used to derive,
by means of the $\derive$ algorithm, 
{\em restricted decryption keys} that can be used to decrypt ciphertexts.
Restricted decryption keys are also associated with a pair $(\y,d)$ where 
$\y$ is an {\em attribute vector} of length $\ell$ over the alphabet 
$\tilde\Sigma=\Sigma\cup\{\star\}$ and $d$ is an integer called the 
{\em distance}.
Keys and ciphertexts interact in the following way.
For vectors $\x\in\Sigma^\ell$ and $\y\in\tilde\Sigma^\ell$ and 
distances $(d',d)$,
define the value of the predicate $\Match((\x,d'),(\y,d))$ to be true iff
for each $i\in[\ell]$ it holds that $y_i=\star$ or $x_i=y_i$ and 
$d\leq d'$. 
In other words, the vectors must match for all positions $i$ 
in which $\y$ is not $\star$ and the ciphertext must have 
a larger distance than the key's. 
The decryption algorithm $\dec$ takes as input 
a ciphertext $\ct$ and a restricted decryption key $\key$ and, if 
the attributes, $(\x,d')$, of $\ct$ and, 
$(\y,d)$ of $\key$ satisfy the $\Match$ predicate
then the plaintext $M$ associated with $\ct$ is returned. If the attribute
vectors do no match, then $\dec$ fails and returns $\perp$.
We stress that the decryption algorithm is oblivious to the attributes of
$\x$ and $\y$ associated with ciphertext and key, respectively, 
which need not to be available in clear to the decryption algorithm.
We observe that 
the master secret key $\msk$
can be seen as the restricted secret key associated with the all-$\star$
attribute vector and distance $0$.

In addition to the above algorithm, a {\primAcr} scheme includes
a key delegation algorithm $\delegate$ that takes as input a 
restricted decryption key for $(\y,d)$ and  an integer $d'>d$ and 
returns a key for pair $(\y,d')$.
We stress that the $\delegate$ algorithm 
need not to know the attribute vector $\y$ associated with 
the key $\key$ being delegated.

Finally, the \revoke\ algorithm is used to revoke encryption keys.
More specifically,  
the \revoke\ algorithm takes as input an encryption key $\pk'$ with
the associated master secret key $\msk'$ and 
a sequence of {\em old} ciphertexts $\ct_1',\ct_2',\ldots$ 
and {\em old} restricted decryption keys $\key_1',\key_2',\ldots,$.
The \revoke\ algorithm returns
a new pair $(\pk^N,\msk^N)$ of encryption and {\em new}
ciphertexts $\ct_1^N,\ct_2^N,\ldots$ and 
new restricted decryption keys $\key_1^N,\key_2^N,\ldots,$.
The crucial property of the {\revoke} algorithm is that 
the new ciphertexts encrypt the same plaintexts with respect to the 
new key as the old ciphertexts did with respect to the old key and,
similarly, the new 
restricted decryption keys have, with respect to the new encryption and 
master secret key, the same attribute that the old keys had with respect
to the old encryption and master secret key.
A trivial way to obtain revocation
is to generate a new pair of encryption and master secret key
using  the $\kg$ algorithm and then to decrypt and re-encrypt the 
ciphertexts with the new public key and to generate the new 
restricted decryption keys using the {\delegate} algorithm.
In our construction of a {\primAcr}, we will use  a much more efficient 
algorithm.


\subsection{Security properties of \primAcr}
In this section we describe the security guarantee that are provided by 
a secure implementation of a \primAcr\ scheme.
As a first security property,
we require that a ciphertext computed with the encryption algorithm
of a \primAcr\  scheme hides the plaintext to anyone that does not have 
the appropriate key. Specifically, suppose that user $A$ encrypts 
plaintext $M$ using a secure \primAcr\ scheme and specifying 
attribute $(\x,d)$ and obtains ciphertext $\ct$.
Then suppose that user $B$ requests and obtains 
restricted decryption keys relative to attribute $(\y,d')$. 
Then we require that $B$ is able to decrypt $\ct$ and thus recover $M$
if and only if $\match((\x,d),(\y,d'))={\tt true}$. 
In addition to protecting the plaintext $M$, we also require that a 
ciphertext does not leak any information about the attribute $\x$ than
what can be obtained from the attribute $\y$.
In other words, $B$ might be able to check if 
$\match((\x,d),(\y,d'))={\tt true}$. If this is the case then $B$ knows
that $\x$ and $\y$ agree in all positions $i$ in which  $y_i\ne\star$
but should have no information about $x_i$ for all positions 
if which $y_i=\star$. 
On the other hand, if 
$\match((\x,d),(\y,d'))={\tt false}$
then $B$ knows that there is at least one
 position $i$ in which $y_i\ne\star$ and 
$x_i\ne y_i$; but $B$ should not know where the mismatch occurs and 
if it occurs for one position or for more than one position.

\ignore{
\subsection{Using {\primAcr} to codify social relationship}
In our OSN a user can perform two basic actions:
to establish a unidirectional link with 
another user; and to publish resources. 
We do not assume that all links and all resources published by a user
belong to the same social network but rather we allow each user 
to specify for each resource and for each
relationship a set of attributes 
(that identifies the social sub-network to which they belong).
Specifically, resources will be labeled with pairs $(\x,d)$ 
where $\x$ is a vector of some pre-specified length $\ell$ 
and $d$ is an integer
and links with pairs $(\y,d')$ where $\y$ is also a vector of length $\ell$ 
and $d'$ is an integer.
If user Alice publishes a resource specifying $(\x,d)$, then 
user Bob sharing a link of type $(\y,d')$ with Alice can 
access the resource if and only if $\match((\x,d),(\y,d'))={\sf true}$.

\paragraph{An example.}
User Alice decides to classify her published resources
according to the following $\ell=2$ categories: 
{\sf Friends} and {\sf coWorkers}.
Alice has friends from high school, college and the music club or neighbors;
similarly co-workers can be partitioned according to the $3$ 
different projects Alice is working on. 
Thus a resource is labeled  with a vector of length $2$ over 
$\{0,1,2,3,4\}$. Notice that since Alice is only involved in $3$ projects,
no resource will ever be labeled with a vector having $4$ as second component.
In addition for each resource Alice specifies a maximum distance at which
the resource can be accessed.
Obviously any such classification is likely to be partial and to change over
time; as we shall see, our system is flexible enough to allow 
Alice to dynamically and efficiently 
change the classification of her links and resources.
Now suppose Alice wants to publish an announcement regarding a concert to
all of her friends from the music club; then she can specify attribute 
vector $\langle3,0\rangle$. Here 
the first attribute (that is, $3$) specifies the category
of friends to which it is addressed ($3$ stands for friends from
the music club) and 
the second attribute specifies the category of co-workers 
(in this case Alice does not want co-workers to receive the announcement about the concert from her and thus specifies $0$). 
In addition Alice can specify a maximum distance $d$ she wants are announcement to be propagated. For example, specifying 
$(\langle3,0\rangle,1)$ makes the concert announcement accessible 
only to friends from the music club; instead 
$(\langle3,0\rangle,2)$ makes the concert announcement accessible
to friends from the music club and to their friends.

Alice can specify attributes also for relationships by giving a pair
$(\y,d)$ where $\y$ is a vector of length $2$ over 
$\tilde\Sigma=\{0,1,2,3,4,\star\}$.
For example,
Alice's link to Bob, a senior colleague of Alice's working on Project1,
could be be labeled with the pair $(\langle 0,1\rangle,1)$.  
The distance in this case can be used to assign weights to links.
For example, the link to Carol, a junior colleague of Alice's working
on Project1, could be labeled with the pair $(\langle 0,1\rangle,2)$.  
In this way a document addressed to senior personnel of project $1$ 
could be labeled with $(\langle 0,1\rangle,1)$.
In classifying links a user is also allowed to use wild cards. 
For example, Alice's supervisor David will be linked to Alice with a link
labeled $(\langle 0,\star\rangle,1)$ which gives David access
to documents from all projects Alice is working on.

As we have seen, 
resources are tagged with an attribute vector and a distance 
encoding the set of users and the distance in the social 
up to which owner allows the resource to be accessed. 
Going back to our previous example of Alice publishing a 
concert announcement with attribute 
$(\langle3,0\rangle,2)$, we make two remarks.
First of all, the distance restriction does not mean that the 
announcement cannot be made available at distance greater than $2$ 
from Alice. 
Indeed, a friend of Alice's can read the announcement and 
re-post it herself.
The distance restriction only guarantees that if the announcement is 
read at distance greater than $2$ from Alice, then it cannot be linked 
to Alice.
As a second remark, 
in our OSN, the attributes to links and to resources are not to be taken
as recommendations. For example, by tagging the 
concert announcement with  
attribute $\langle 3,0\rangle$,
Alice does not mean that co-workers should not 
look at the announcement and that she relies on the the co-workers' 
honesty in not looking at the announcement 
(experience shows that such a labeling  would be too strong of a temptation
to resist for most people). 
Rather the resource will not be made available to co-workers  and 
this will not be enforced by the central manager of the OSN but rather 
we will provide Alice with a publishing procedure that will 
cryptographically
enforce Alice's decision of the set of users with which we wants to share 
the announcement.
The relevance of {\primAcr} to our proposal of OSN should be a clear
at this point.
}

\section{Construction of a secure \primAcr}
\label{sec:primitives}
First, we present HVE and HIBE and then we show how they can be used to construct a \primAcr\ scheme.

\subsection{Hidden Vector Encryption}
A Hidden Vector Encryption (HVE for short) scheme  consists of four algorithms $(\HKg,\HEnc,\HDec,$ $\HDerive)$.
The syntax and semantics of a HVE scheme is very similar to  that of a \primAcr\ scheme. For sake of completeness we include it.
The key generation algorithm $\HKg$ returns an {\em encryption key} $\pk$ 
and a {\em master secret key} $\hmsk$.  
The encryption algorithm $\HEnc$ takes as input 
the encryption key $\pk$, a plaintext $M$ and 
{\em ciphertext type} $\x$ consisting of an {\em attribute vector} $\x$ of length $\ell$ 
over a fixed alphabet $\Sigma$.
The master secret key $\hmsk$ can be used to derive,
by means of the $\HDerive$ algorithm, 
{\em restricted decryption keys} that can be used to decrypt ciphertexts.
Restricted decryption keys are also associated with a vector $\y$ where 
$\y$ is an {\em attribute vector} of length $\ell$ over the alphabet 
$\tilde\Sigma=\Sigma\cup\{\star\}$.
Keys and ciphertexts of HVE interact in the following way.
For vectors $\x\in\Sigma^\ell$ and $\y\in\tilde\Sigma^\ell$,
define the value of the predicate $\HMatch(\x,\y)$ to be true iff
for each $i\in[\ell]$ it holds that $y_i=\star$ or $x_i=y_i$. 
In other words, the vectors must match for all positions $i$ 
in which $\y$ is not $\star$.
The decryption algorithm $\HDec$ takes as input 
a ciphertext $\ct$ and a restricted decryption $\key_H$ and, if 
the attributes, $\x$, of $\ct$ and, 
$\y$ of $\key$ satisfy the $\HMatch$ predicate
then the plaintext $M$ associated with $\ct$ is returned. If the attribute
vectors do no match, then $\HDec$ fails and returns $\perp$.
We stress that the decryption algorithm is oblivious to the attributes of
$\x$ and $\y$ associated with ciphertext and key, respectively, 
which need not to be available in clear to the decryption algorithm.
We observe that 
the master secret key $\hmsk$
can be seen as the restricted secret key associated with the all-$\star$
attribute vector.

\subsubsection{Security properties of HVE}
In this section we describe the security guarantee that are provided by 
a secure implementation of a HVE\ scheme.
As a first security property,
we require that a ciphertext computed with the encryption algorithm
of a HVE  scheme hides the the plaintext to anyone that does not have 
the appropriate key. Specifically, suppose that user $A$ encrypts 
plaintext $M$ using a secure HVE scheme and specifying 
attribute $\x$ and obtains ciphertext $\ct$.
Then suppose that user $B$ requests and obtains 
restricted decryption keys relative to attribute $\y$. 
Then we require that $B$ is able to decrypt $\ct$ and thus recover $M$
if and only if $\HMatch(\x,\y))={\tt true}$. 
In addition to protecting the plaintext $M$, we also require that a 
ciphertext does not leak any information about the attribute $\x$ than
what can be obtained from the attribute $\y$.
In other words, $B$ might be able to check if 
$\HMatch(\x,\y)={\tt true}$. If this is the case then $B$ knows
that $\x$ and $\y$ agree in all positions $i$ in which  $y_i\ne\star$
but should have no information about $x_i$ for all positions 
in which $y_i=\star$. 
On the other hand, if 
$\HMatch(\x,\y)={\tt false}$
then $B$ knows that there is at least one
 position $i$ in which $y_i\ne\star$ and 
$x_i\ne y_i$; but $B$ should not know where the mismatch occurs and 
if it occurs for one position or for more than one position.
For our implementation we use the HVE system of \cite{DBLP:conf/pairing/IovinoP08}.

\subsection{Hierarchical Identity-based Encryption}
A Hierarchical Identity-based Encryption (HIBE for short) \ scheme consists of six algorithms
$(\IKg,\IEnc,\IDec,\IDerive,$ $\IDelegate,\IRevoke)$.
The key generation algorithm $\IKg$ returns an {\em encryption key} $\pk_I$ 
and a {\em master secret key} $\imsk$.  
The encryption algorithm $\IEnc$ takes as input 
the encryption key $\pk_I$, a plaintext $M$ and 
{\em ciphertext type} $\vec {id}$ consisting of a vector of length $\le \ell$ over the alphabet $\Sigma$.
The master secret key $\imsk$ can be used to derive,
by means of the $\IDerive$ algorithm, 
{\em restricted decryption keys} that can be used to decrypt ciphertexts.
Restricted decryption keys are also associated with vectors $\vec {id}$ of length $\ell$ over the alphabet $\Sigma$.
Keys and ciphertexts of a HIBE interact in the following way.
For vectors $\vec {id}\in\Sigma^{m}$ and $\vec {id'}\in\Sigma^{n}$, $m\le n\le\ell$, 
define the value of the predicate $\prefix(\vec {id},\vec {id'})$ to be true iff
for each $i\in[m]$ it holds that $id_i=id_i'$. 
In other words, the $\prefix$ predicate checks if vector $\vec {id}$ is a prefix of $\vec {id'}$.
The decryption algorithm $\IDec$ takes as input 
a ciphertext $\ct$ that encrypt a plaintext $M$ and a restricted decryption key $\key_I$ and, if 
the vectors, $\vec {id}$, of $\ct$ and, 
$\vec {id'}$ of $\key_I$ satisfy the $\prefix$ predicate
then $M$ is returned, otherwise $\IDec$ fails and returns $\perp$.
We observe that 
the master secret key $\imsk$
can be seen as the restricted secret key associated with the length $0$ vector.
In addition to the above algorithm, a HIBE scheme includes
a key delegation algorithm $\IDelegate$ that takes as input a 
restricted decryption key for $\vec {id}$ and  another  vector $\vec {id'}$ such that $\prefix(\vec {id},\vec {id'})$ holds and returns a key for the new vector $\vec {id'}$.

Finally, the $\IRevoke$ algorithm used for revocation has the same properties than the one for $\primAcr$.

\subsubsection{Our HIBE system}
Our implementation uses the same HIBE of Boneh, Boyen and Goh \cite{DBLP:conf/eurocrypt/BonehBG05} (BBG for short) that is efficient and offers constant-size ciphertexts.
In addition our HIBE system include a revocation procedure not present in the original scheme.
In the following, we assume that the reader is familiar with bilinear groups.
The $\IRevoke$ procedure works as follows.
Recall that in the BBG system the secret is an element $g_2^\alpha$ in the base group and the public key contains $g,g_1=g^\alpha,g_2$ so all these elements are given as input to the revocation procedure.
The revocation procedure re-randomizes $g_2^{\alpha}$ by setting the new $g_2^{\alpha'}$  to be equal to $g_2^{\alpha+\beta}$ for a random group element $g_2^{\beta}$.
Then, the procedure re-randomizes the only term in the decryption key where the old $\alpha$ appeared.
Indeed, in the BBG system the first group element of the decryption keys contains the term $g_2^{\alpha}$ so the procedure can multiply such first group element by $g_2^{\beta}$ to obtain a new key consistent with the new secret $\alpha'=
\alpha+\beta$. Analogously, the public key $g^{\alpha}$ is multiplied by $g^{\beta}$ to obtain a new public key with the consistent distribution.
Finally, notice that the only group element in the ciphertext containing $\alpha$ is 
$\Omega=e(g_1,g_2)^s=\e(g,g_2)^{\alpha s}$ and notice that the ciphertext also contains the group element $g^s$ so
the re-randomization can be obtained by computing $\Omega'=\e(g^s,g_2)^\beta=\e(g,g_2)^{\beta s}$ and multiplying $\Omega$ by $\Omega'$ to obtain the new value $\e(g^{\alpha'},g_2)^{s}$ with the right distribution.
It is easy to see that the old keys can not decrypt the new ciphertexts anymore and viceversa.
We stress that the above procedure is efficient because it takes time independent of $\ell$, the maximum length of the identity vectors.
\subsubsection{Security properties of HIBE}
We require that a ciphertext computed with the encryption algorithm
of a HIBE  scheme hides the the plaintext to anyone that does not have 
the appropriate key. Specifically, suppose that user $A$ encrypts 
plaintext $M$ using a secure HIBE scheme and specifying 
vector $\vec {id}$ and obtains ciphertext $\ct$.
Then suppose that user $B$ requests and obtains 
restricted decryption keys relative to vector $\vec {id'}$. 
Then we require that $B$ is able to decrypt $\ct$ and thus recover $M$
if and only if $\prefix(\vec {id},\vec {id'})={\tt true}$.

\subsection{Implementation of a secure \primAcr\ scheme using HVE and HIBE}
We now show how to construct a secure \primAcr\ scheme by using a secure HVE and HIBE schemes.
The $\kg$ algorithm calls the algorithms $\HKg$ and $\IKg$ to obtain the pairs $(\hmsk,\pk)$ and $(\imsk,\ipk)$ and sets $\msk=((\hmsk,\imsk)$ and $\pk=(\hpk,\ipk)$.

The encryption algorithm $\enc$ take as input a message $M$ and a pair $(\x,d)$, and proceeds as follows.
It encrypts $M$ and $\x$ by using the HVE encryption $\HEnc$ with public key $\pk$ to produce ciphertext $\ct_H$.
Finally it encrypts by mean of $\IEnc$ and public key $\pk_I$ the message $\ct_H$ with identity $\vec {id}=1^d$, that is the distance $d$ codified in unary.
The encryption $\enc$ returns the output $\IEnc$.

The $\derive$ algorithm takes as input the master secret key $\msk$ and a pair $(\y,d)$ and proceeds as follows.
It calls $\HKg(\hmsk,\y)$ to obtain $\key_H$.
It calls $\IKg(\hmsk,\vec {id})$, where $\vec {id}=1^d$ is a vector that codifies $d$ in unary,  to obtain $\key_I$.
The key output by $\derive$ is set to the pair $\key_H$ and $\key_I$.

The decryption algorithm $\dec$ takes as input a key $\key=(\key_H,\key_I)$ and a ciphertext $\ct$ that encrypts the plaintext $M$ and works as follows.
It uses the decryption algorithm $\IDec$ of HIBE with key $\key_I$ to decrypt $\ct$ (recall that $\ct$ is a ciphertext type of HIBE) and obtains $\ct_H$.
Now use the procedure $\HDec$ of HVE with input $\ct_H$ and key $\key_H$ to obtain the plaintext $M$.

The $\delegate$ procedure takes as input a key $\key=(\key_H,\key_I)$ for distance $d$ and a new distance $d'>d$ and works as follows.
It computes $\key_I'=\IDerive(\key_I,1^{d'})$, where $1^{d'}$ is an encoding of $d'$ in unary, and returns the new key pair $\key'=(\key_H,\key_I')$.

The $\revoke$ procedure calls the revocation procedure for the HIBE system.

It is possible to prove that the so constructed \primAcr\ scheme is correct and secure assuming the correctness and security of the underlying HVE and HIBE schemes.

\section{Access policy enforcement}
\label{sec:enfo}
We now describe the protocols required to perform the functionalists introduced in Section \ref{sec:framework}.
We refer to the terminology introduced in Section \ref{sec:framework}. 
Moreover, in the following we assume that each user of the social network has already executed the requested cryptographic setup (parameters generation, etc.).

Furthermore, in the following we assume that the user $u$ executing the protocols for protecting her/his resources, has defined the set
$\mathcal{C}_u = \left\{cond_1, \ldots, cond_n\right\}$ of the conditions which she/he is going to use in the definition of the policies protecting the sensitive resources owned by her/him.
We recall from Section~\ref{sec:lang} that a condition is an expression of the form $res.attr\,pred\,val$ where $res$ is a resource name, $attr$ is an attribute name from the ones over which the resource $res$ is defined, $val$ is a valid value from the corresponding domain of the attribute $attr$ and $pred$ is a binary predicate from $\{=, \leq, <, >, \geq,\not =\}$.

\remove{
\begin{definition}[Policy vector]
Given an access rule for resource $res$  $ar^{res}=cond_1, \ldots, cond_k$, the \emph{policy vector} associated to $pol$ is a vector $\vec{v}_{ar^{res}} = [v_1, \ldots, v_n]$ where $v_i = 1$ if the condition $c_i \in \mathcal{C}_u$ is stated in $pol$, $0$ otherwise. We omit super/subscripts when they are evident from context.
\end{definition}  
}

\subsection{Enrollment}
\label{sec:enrollment}
We assume that the OSN has fixed a \primAcr\ scheme and a symmetric key encryption 
scheme \ske.
Upon enrolling into the OSN, a user $A$ generates a pair 
$(\pk,\msk)$ of encryption and master secret key for \primAcr\ by running 
algorithm \kg\ on input the maximum distance $d_{\rm max}$.
The encryption key is published in a publicly-accessible 
repository whereas the master secret key is kept by $A$ in a private 
repository.

\subsection{Resource publication and access}
\label{sec:publication}
The encrypted version of the sensitive resource is published on an untrusted repository. 
As such, we cannot assume the repository enforces the access control policy associated with 
the resource. As we have explained in the sections above, the access control policy is encoded 
by means of a {\em policy pair} $(\x,d)$ consisting of an 
attribute vector $\x$ and of a maximum distance $d$ . The policy pair is 
used by the encryption procedure of {\primAcr}.
More precisely, 
the publication of a resource $R$ protected by access policy $\pol$ 
(and therefore by the corresponding policy pair $(\x_{\pol},d_\pol)$)
is carried out by performing the following three steps:
\begin{enumerate}
\item User $A$ generates a random secret key $\sk_R$ for symmetric encryption scheme \ske\
by running algorithm $\skg$;
\item resource $R$ is encrypted with key $\sk$ by running algorithm $\senc$ on input $\sk_R$ and $R$ obtaining the 
{\em permanent ciphertext} $\ct_R$ for resource $R$; 
\item secret key $\sk$ is encrypted using encryption key $\pk$ with 
policy pair $(\x_{\pol},d_\pol)$  and algorithm \enc\ to obtain the 
{\em revocable ciphertext} $\tilde\ct_R$ for resource $R$.
\end{enumerate}
The pair of permanent ciphertext and revocable ciphertext 
$(\ct_R,\tilde\ct_R)$ are sent to the public repository.

A user that wants to access a resource $R$ with policy pair $(\x,d)$ using a
restricted decryption key with policy pair $(\y,d')$ such that 
$\match((\x,d),(\y,d'))={\tt true}$ first decrypts the revocable ciphertext
$\tilde\ct_R$ using algorithm  \dec\ and thus obtaining secret key $\sk_R$
and then uses $\sk_R$ as input to $\sdec$ to decrypt the permanent 
ciphertext $\ct_R$.

\subsection{Link creation and revocation}
\label{sec:link}
In this section we describe how user $A$ can establish link to user $B$
and assign to the link policy pair $(\y,d')$ where $\y$ is a vector over 
$\tilde\Sigma$ and $d'>0$ is an integer.
User $A$ runs algorithm \derive\ on input the master secret key \msk\ 
computed at  enrollment and pair $(\y,d)$ to obtain a restricted decryption
key $\key$ that is sent to user $B$ along with encryption key $\pk$.
The restricted decryption key $\key_{A\leftarrow B}$ will be used by $B$ to access 
all resources published by $A$. 
In addition, 
upon receiving the pair $(\pk,\key)$ user $B$ propagates the restricted 
decryption key to all users user $C$ for which $B$ has established a link.
Specifically, suppose $B$ has established a link with policy pair 
$(\z,e)$. Then, using the \delegate\ algorithm $B$ derives a key with 
policy pair $(\y,d+e)$ and sends it to $C$. $C$ does the same with all her
neighbors until maximum distance $d_{\rm max}$ is reached.

When user $A$ wants to revoke her link with user $B$, 
she executes the \revoke\ algorithm on input the 
encryption key $\pk$, the master secret key $\msk$, 
the revocable ciphertexts $\tilde\ct_R$, 
for all published resources $R$ and 
the restricted decryption keys $\key_{A\leftarrow C}$ for all $C\ne B$ for 
which $A$ has established a link. 
As output, $A$ obtains a new public key $\pk^N$,  a new master secret key 
$\msk^N$, new revocable ciphertexts $\tilde\ct_R^N$, for each resource $R$
and new 
restricted decryption keys $\key_{A\leftarrow C}^N$.

Finally, $A$ replace $\pk$ with $\pk^N$ in the public repository, 
$\msk$ with $\msk^N$ in the private repository, replaces the revocable 
ciphertexts $\tilde\ct_R$ with $\tilde\ct_R^N$ in the public repositories
and sends the new 
restricted decryption key $\key_{A\leftarrow C}^N$ to each user $C$.


\ignore{
\subsection{Link creation}
\label{sec:link}
When a user $u'$ contacts the user $u$ in order to establish a friendship link with her/him, $u'$ sends to $u$ (the values of) his/her attributes
\footnote{or, at least, those values that $u'$ believes can be safely disclosed and can be used by $u$ to give access to her/his resources}, as well.
Upon receiving the attributes of $u'$, $u$ checks whether the received values satisfy any of the access policies that she/he has previously defined.
For each resource that $u'$ is allowed to access according to the policies, $u$ creates a vector $\vec{y} = [y_1,\ldots,y_k]$, with $k \geq l+1$, representing the attributes identified by $u$ in $u'$ and the distance allowed by the key which will generated in the follows.
As for the vector $\vec{x}$ used in the resource publication protocol, the components of $\vec{y}$ are partitioned in two parts.
The first part, composed by $y_i$ with $i \in [1, l]$, represents the distance assigned to the key and, implicitly, the delegation capability of the user receiving the key created by means of $\vec{y}$ (we will discuss delegation in Section \ref{sec:propagation}).
The second part, 
which means the elements $y_j$ with $j \in [l+1, k-1]$ are used to represent the conditions verified by the attributes provided by $u'$.
Hence, $y_j = 1$ if and only if $c_{j-l}$ holds true with respect to the attributes of $u'$, $0$ otherwise.
Moreover, $y_k = x_k$ for $x_k$ being a random value generated during the resource publication protocol.

More precisely, the first $l$ entries of $\vec{y}$, which we recall represents the distance within which the other are allowed to access the resource $r$, are assigned as follows, given a maximum allowed distance $d$:
\begin{itemize}
\item $y_1 = 1$ identifying that $u'$ is at least at distance one from the owner of the resource;
\item $y_i = 0$ for $i \in [d+1, l]$ for a given distance $d$ defined by the user $u$ and encoded in $\vec{x}$, such reflects the maximum distance encoded in $\vec{x}$;
\item $y_j = \ast$ for $j \in [2, m]$, where $m \leq d$ is the maximum delegation distance allowed to $u'$ with respect to her/his attributes and
\item $y_h = 1$ for $h \in [m+1, d]$, assignments made to match the distance encoded in $\vec{x}$ but not allowing $u'$ to further propagate the key obtained to users at more than $m$ hops from $u$.
\end{itemize}
Finally, the user $u$ computes $K_{u'} = \textsf{HHVE.KeyGen}(MSK, \vec{y})$, where $MSK$ is the master secret key for HHVE computed during the setup phase, and sends $K_{u'}$ to $u'$.
}

\ignore{
\subsection{Key delegation}
\label{sec:propagation}
As mentioned in Section~\ref{sec:intro} the users of the OSN are entitled to propagate the secret keys of their respective key rings according to a delegation protocol.



Note that according to the proposed framework, a user $u'$ is allowed to delegate a key $K_{u'}$ obtained by $u$ iff the attribute vector $\vec{y}$ associated to $K_{u'}$ contains $\ast$ in the first $l$ positions.
Thus, the delegation of a key $K_{u'}$ owned by a user $u'$ to a user $u''$ is performed executing the following protocol:

\begin{enumerate}
\item User $u'$ asks 
 to $u''$ her/his attributes in order to verify $u''$ satisfies some of the access control policies defined by $u'$ to limit the propagation of the keys in her/his keyring.
\item If $u''$ satisfies the access control policy protecting $K_{u'}$ and if the vector $\vec{y}$ associated to the key contains at least a $\ast$ in the fist $l$ positions, then $u'$ compute a vector $\vec{z}$ where
$z_i = y_i$ for $i \in [1, d'-1] \cup [d'+1,k]$ and $d' = min(\{j: j \in [1, l] \wedge y_j = \ast\})$.
Moreover, $z_{d'} = 1$.
\item User $u'$ computes the delegation key $K'_{u'} =\textsf{HHVE.Delegate}(K_{u'},\vec{y},\vec{z})$ and sends it to $u''$.
\end{enumerate}
}

\ignore{
\subsection{Resource access}
\label{sec:access}

When a user $u'$ wants to access a resource $r$ published on the repository by another user $u$, $u'$ retrieves from the repository the tuple $(\hat{R}, \hat{SK})$.
We recall that $\hat{R}$ is the resource $r$ encrypted by means of a symmetric encryption scheme SKE using the unique key $SK$ and that $\hat{SK}$ is the corresponding key encrypted by means of the attribute based public encryption scheme HHVE.
Assuming that the user $u'$ owns a keyring $\mathcal{K}_{u'} = \{K_1, \ldots, K_s\}$, the access of the resource is performed according to the following steps.
$u'$ verifies if $\textsf{HHVE.Check}(\hat{SK}, K_i)$ holds, for $i \in [1, s]$.
If there exists a key $K_j$ which is able to decrypt $\hat{SK}$ then $u'$ retrieve $SK$ and, after that, computes $r = \textsf{SKE.D}(\hat{R}, SK)$.
On the other hand, if $u'$ does not own a key able to decrypt $\hat{SK}$, then she/he asks to the repository the identity of the user $u$, the original owner of the resource.
With such information, $u'$ may try to establish a friendship link with her/him satisfying the access control policies defined by $u$ (see Section~\ref{sec:link}).

}

\ignore{
\subsection{Key revocation}
\label{sec:revocation}




When a user $u$ wants to revoke access on a given resource $R$ to a given user $u'$ -- and subsequently to all the users $u'_1, \ldots, u'_f$ who have been authorized by $u'$ by means of the delegation property of HHVE -- , she/he must revoke the key $K_{u'}$ specifically created for $u'$.
Given the attribute vector $\vec{x}$ used during the generation of $K_{u'}$ and given the revocation of all the keys associated with $\vec{x}$ is performed generating a new $SK$ associated with $R$.

After that, a new vector $\vec{x'}$ is generated as $\forall i \in {[1, k-1]}$ $x'_i = x_i$ and $x'_k = r'$ where $r'$ is a new randomly generated value.
The random value $r'$ and $\vec{x'}$ are used to compute a new $\hat{SK}'$ which will substitute $\hat{SK}$ on the remote repository. 

As a result, all the keys -- both propagated to direct contacts and delegated by direct contacts to farer nodes of the OSN -- are revoked.
Thus, it is required that $u$ sends a newly generated key $K_v$, with $v \not = u'$ to all users who are her/his contacts and who are still allowed to access $R$.

%

}

\section{Experimental results}
\label{sec:exp}
In order to evaluate the feasibility of the proposed access control model and of the related enforcement protocols, we developed a prototype application providing the features described in the previous sections.
The prototype has been developed in Java 6 using the jPBC~(\cite{jpbc,DeIo11}) and the BouncyCastle~\cite{bouncyCastle} cryptographic libraries and it is deployed as a Facebook application.

\begin{figure}[htb!]
\centering
\includegraphics[width=.9\columnwidth]{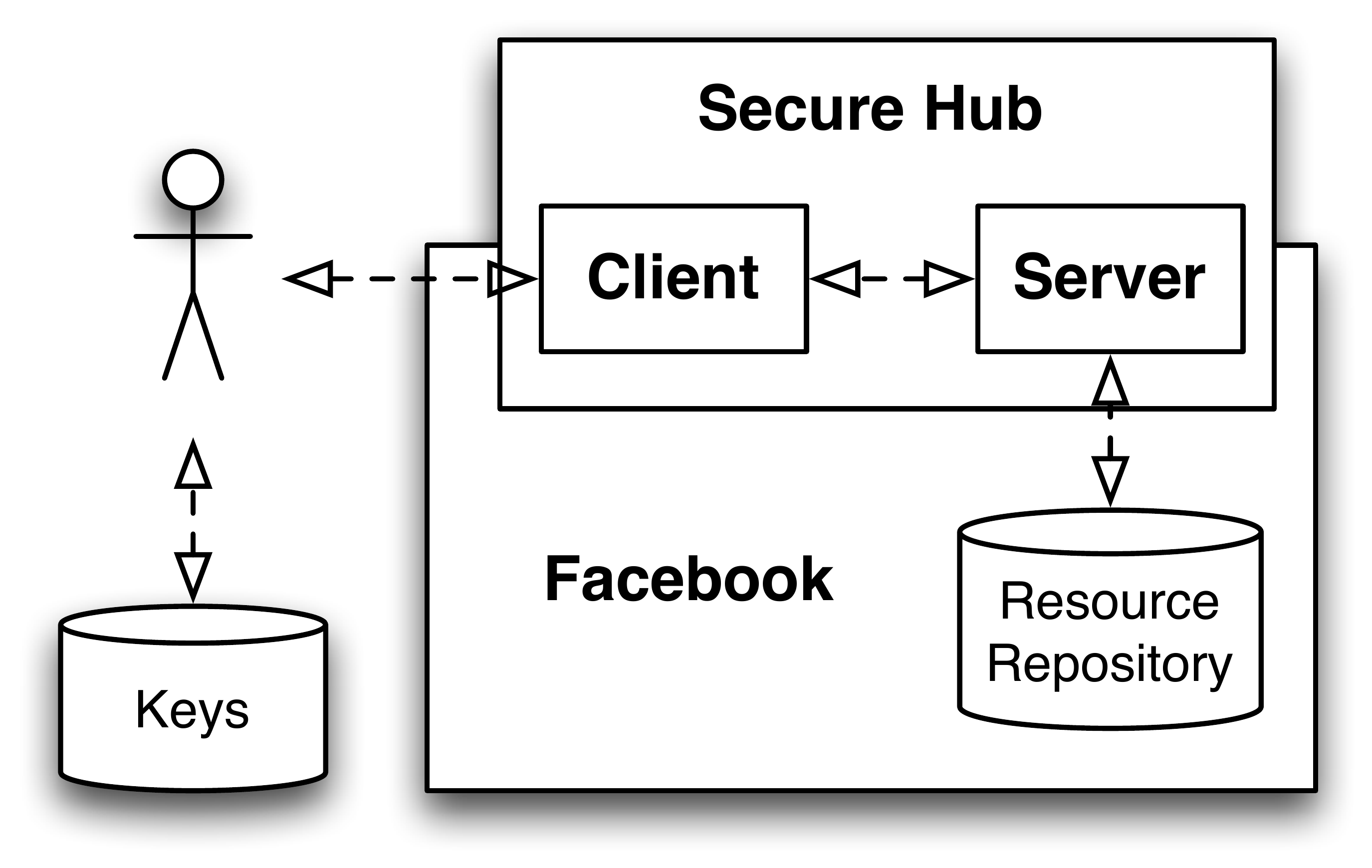}
\caption{The architecture of the prototype}
\label{fig:architecture}
\end{figure}

The application architecture follows the client-server paradigm, as shown in Figure~\ref{fig:architecture}.
A user interacts through a web interface with the client module, which is in charge of generating the user keys and of the encryption/decryption of the resources.
The user is allowed to define the access policy to be associated to the resource which she/he is currently uploading on Facebook.
Upon the upload in the client module of the resource, such module execute the steps described in 
paragraph on resource publication 
and it sends the encrypted resource to the server module, which is in charge of storing it in the resource repository.

The client module is also in charge of retrieving the list of friends of the executing user and, after a check against their attributes, generates the search keys required to access the uploaded resources.
Further, upon receiving a search key, the client module propagates such key to the owner's Facebook friends, following the steps described for creating a link.

The purpose of the server module is to store the encrypted resources, alongside with metadata such as the resource names and the references to owners, and to provide a common interface to search and retrieve them.
Note that the decryption of the resource is performed on the client side.
Thus, no cryptographic material -- such as keys -- is available to the server module, other than the encrypted resources. 

\begin{figure}[htb]
\centering
\subfigure[Resource dimension in KiB]{
   \includegraphics[width=.5\columnwidth] {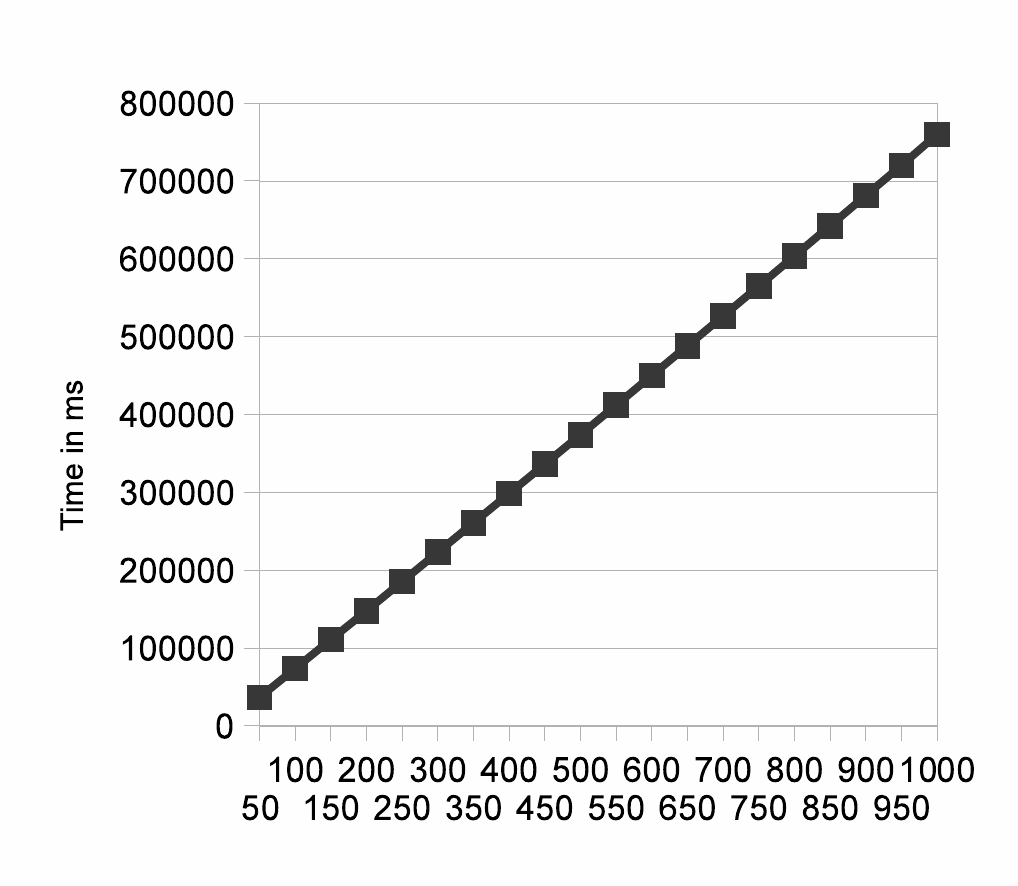}
   \label{fig:resultsRes}
 }
\subfigure[Policy vector length]{
   \includegraphics[width=.4\columnwidth] {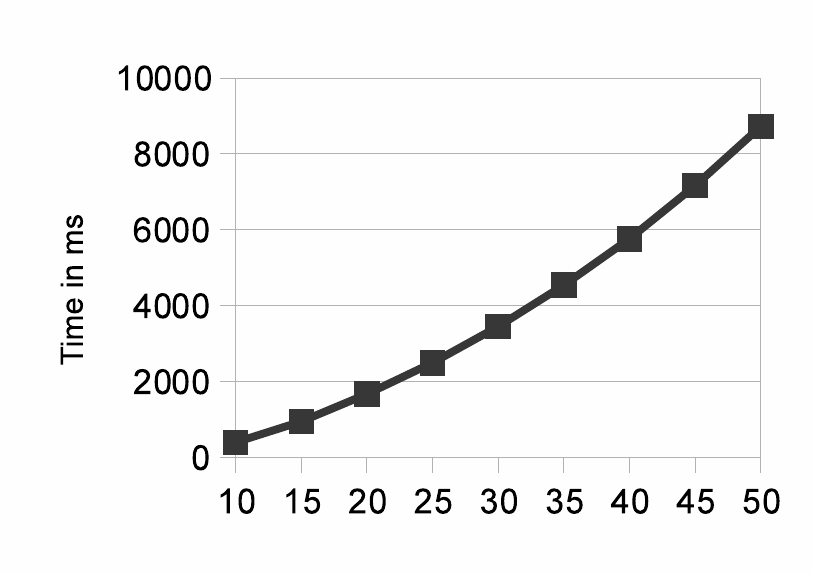}
   \label{fig:resultsVect}
 }

 
 \caption{Experimental results}
 \label{fig:results}
\end{figure}

We performed two series of experiments to evaluate the performances of the 
\primAcr\ scheme.
More precisely, we tested the encryption algorithm varying the size of the shared resource and the size of the vector encoding the access control policy.
The obtained results show that the performances are linear with respect of both the parameters, as shown in Figure~\ref{fig:results}.

\section{Related Work}
\label{sec:relwork}

Access control in OSNs is a relatively new research area in which -- nevertheless -- several access control models and related mechanisms have been presented, aiming to overcome the restrictions of the mechanisms provided by current OSNs (see \cite{CF10} for a survey).
One of the common characteristics of almost all the defined access control models is that access control is relationship-based, that is, authorized users are denoted on the basis of constraints on the relationships the requester should have with other network users. In particular, such constraints refer to the depth of the relationship and/or its type.
As an example, in \cite{DBLP:journals/tissec/CarminatiFP09} the authors present a framework for OSN access control that allows the definition of suitable policies with respect to the relationships occurring among the users and the corresponding associated trust values.

In the recent years,several work have been done to the definition of cryptographic schemes and protocols to store sensitive data over untrusted storage.
As an example, in \cite{DBLP:conf/ndss/GohSMB03} the authors propose the deployment of a cryptographic layer over existing network file systems to provide privacy and confidentiality of the stored files. Their approach uses asymmetric cryptography to allow the sharing of the files among different users.
In \cite{DBLP:conf/storagess/NaorSW05} it is presented an alternative approach which avoid the use of asymmetric cryptography in order to achieve better performance, with respect to computational time, however maintaining the flexibility  of the key management typical of asymmetric scheme.

In 
\cite{DBLP:conf/ndss/TraynorBEM08} is presented a mechanism to perform access control to \emph{IPtv} streaming data using an ABE scheme. In particular, the authors evaluate the impact of the deployment of ABE schemes in a system requiring massive scalability.
In \cite{DBLP:journals/jcs/PirrettiTMW10} the authors present  an information management architecture based on an ABE scheme. Similarly to our approach, the architecture proposed uses the attribute-based crypto machinery in order to enforce access control policies, which are defined upon verifiable users' attributes.

The introduction of cryptographic primitives in OSN in order to enhance the privacy and the security of shared data is an active research topic.
In \cite{DBLP:conf/sigcomm/AndersonDBS09}, the authors propose an alternative architecture for a distributed OSN, in which a four-layer client-server deploys cryptographic primitives and sandboxed environments to safely share and access information.

\paragraph{The Persona framework}
The work with which we share more common traits is 
Persona \cite{DBLP:conf/sigcomm/BadenBSBS09}.
In it, the authors present an access control framework for OSN using an ABE schema, namely the schema deployed in in the library cpabe~\cite{cpabe}.
In this way, Persona allows users to apply fine-grained policies stating which users view their data.
The main differences between Persona's and our approach 
can be summarized in the following four points:
(i) {\em more expressive access policy definition language.}
Namely, our access policy language allows the definition 
of more comprehensive access control policies. 
First of all, in our policy definition language it is possible 
to express distance constraints and, more importantly, 
link labels can also contain $\star$ entries. 
This means that if 
the label of the link from Alice to Bob has a $\star$ in position $i$, 
then Bob can access resources regardless of the $i$-th attribute.
(ii) {\em Access delegation.} 
That is, our framework allows a user to delegate access to certain resources to her/his contacts, according to the propagation limits defined.
(iii) {\em Efficient access revocation}. 
Our framework allows a user to efficiently revoke all the generated keys which allow access to a given resource, automatically revoking the delegation as well.
Note that in Persona access delegation and revocation are not supported.
(iv) {\em Privacy of the access control policy.} 
Using our framework, 
the policy which protects a resource is known only to the owner.
This is because the policy is encoded by a pair $(\x,d)$ and vector $\x$
is used to encrypt the resource. 
However, we require (and our implementation supports) that the encrypted resources not only hides the resource but also the 
attribute. 

We achieve the four improvements listed above by developing a new 
cryptographic primitives, \primAcr, and by showing that such a more
powerful  primitive can still be implemented in an efficient way.


\section{Conclusions}
\label{sec:conc}
In this work we have introduced a framework allowing users of a OSN to express and enforce access control policies for sharing sensitive information/resources without relying on a possibly untrusted third party, except for the storage of the encrypted information/resource.
Our solution is based on a novel attribute-based encryption scheme and offers several advantages with respect to the existing literature, such as the possibility of expressing private access policies based on the topology of the OSN graph, with efficient revocation.
We have implemented our framework as a Facebook application demonstrating the viability of our approach and showing that we can reach high levels of privacy with reasonable performances.

As further work we are currently working on a fully distributed OSN setting in which we plan to deploy the framework proposed in this work.
Subsequently we plain to perform more extensive experiments and to test other cryptographic primitives.

\bibliographystyle{IEEEtran}
\bibliography{SecSN2}

\end{document}